# Experimental Demonstration of High-Rate Discrete-Modulated Continuous-Variable Quantum Key Distribution System

Yan Pan[1], Heng Wang[1], Yun Shao[1], Yaodi Pi[1], Yang Li[1], Bin Liu[2], Wei Huang[1†], and Bingjie Xu[1*]

[1]*Science and Technology on Communication Security Laboratory, Institute of Southwestern Communication, Chengdu 610041, China*
*\*Corresponding author: huangwei096505@aliyun.com and xbjpku@163.com*
[2]*College of Computer Science, Chongqing University, Chongqing 400044, China;*

A high-rate continuous-variable quantum key distribution (CV-QKD) system based on high-order discrete modulation is experimentally investigated. With the help of the novel system scheme, effective digital signal processing algorithms and advanced analytical security proof method, the transmission results of 5 km, 10 km, 25 km, and 50 km are achieved for the 1 GBaud optimized quantum signals. Correspondingly, the asymptotic secret key rate (SKR) is 288.421 Mbps, 159.395 Mbps, 50.004 Mbps and 7.579 Mbps for discrete Gaussian (DG) 64QAM, and 326.708 Mbps, 179.348 Mbps, 50.623 Mbps and 9.212 Mbps for DG 256QAM. Under the same parameters, the achieved SKRs of DG 256QAM is almost same to ideal Gaussian modulation. In this case, the demonstrated high-rate discrete modulated CV-QKD system has the application potential for high speed security communication under tens of kilometers. © 2022 Optical Society of America

The rise of advanced computing technologies such as quantum computing poses a potential threat to the traditional encryption system. Therefore, exploring reliable technology to achieve a high-security encryption system is meaningful. According to the information security theory proposed by Shannon in 1949, one-time pad is a promising way for the unconditional security encryption scheme [1]. However, the huge key consumption will be a great challenge for traditional key distribution methods. Quantum key distribution (QKD) has brought the dawn of one-time pad technology with its physically provable security and high key rate. Since the BB84 protocol was proposed, the QKD has been mainly developed into two technical routes: discrete-variation (DV) and continuous-variation (CV) [2-5]. Among them, CV-QKD technology has attracted much attention with the advantages of the high key rate and compatibility of optical communication systems, even though the transmission distance of CV-QKD is much shorter than DV-QKD. Early, Gaussian modulated coherent-state (GMCS) protocol is one of the most popular schemes for the CV-QKD system, based on which many high-rate or long-distance experimental demonstrated CV-QKD systems have been reported [6-9]. However, a real Gaussian modulation can never be perfectly implemented since in practice devices such as digital to analog converter (DAC) have finite precision. Meanwhile, the error correction is complex for the Gaussian signal at low SNR.

To solve the above problems, the discrete modulated (DM) CV-QKD protocols are proposed [10]. Unfortunately, owing to its inherent characteristic of non-Gaussian, the security proof is quite difficult. There are various proofs for specific DM CV-QKD protocols ( e.g. two-state and three-state protocols) [11, 12]. Moreover, versatile numerical methods are applied to study the security of DM CV-QKD protocols, but mainly the four-state protocol is investigated [13, 14]. In this case, the complete security proof of the arbitrary discrete modulated protocol that gives a higher key rate is challenging. Recently, an analytical lower bound on the asymptotic secret key rate of CV-QKD with an arbitrary modulation of coherent states was presented, which provides an important tool for experimental works of DM CV-QKD [15]. In the aspect of experimental demonstrations, several groups performed independent studies for local local oscillator (LLO) DM CV-QKD [16-19]. Most of them, for example in Ref. [16-18], utilize the linear channel assumption for calculating the security key rates (SKRs), which is hard to protect against arbitrary collective attacks. In Ref. [19], the authors show experimental results of 67.6 and 66.8 Mbps SKRs based on analytical security proof proposed in Ref. [15]. However, the experimentally demonstrated transmission distance is only 9.5 km, compared to the previous works of LLO CV-QKD, there is more room for improvement in transmission distance.

In this paper, under the security framework of Ref. [15], we experimentally demonstrated a high-rate LLO DM CV-QKD system with various distances based on 1 GBaud discrete Gaussian (DG) 64QAM and 256QAM quantum signals. In the investigated LLO CV-QKD system, polarization and frequency division multiplexing is used to improve the isolation of weak quantum signal and strong reference signal, intermediate frequency detection combined with digital demodulation for $x$ and $p$ quadrature is used to improve the detection efficiency, and polarization diversity receiver and polarization tracking algorithm are used to improve the system stability and flexibility. Here, a real-valued finite-impulse response (FIR) filter is implemented to correct polarization variation, $x$ and $p$ quadrature imbalance compensation, and residual inter-symbol interference. After optimizing the probability shaping and modulation variance by simulations, high asymptotic SKRs are obtained with the help of digital signal processing (DSP) algorithms in the experiment. For DG 64QAM/256QAM, the final experimental asymptotic SKRs are 288.421/326.708 Mbps, 159.395/179.348 Mbps, 50.004/50.623 Mbps and 7.579/9.212 Mbps after transmission over 5 km, 10 km, 25 km and 50km SSMF, respectively. Therefore, the investigated LLO CV-QKD scheme has the application potential for high-rate secure communication under tens of kilometers.

Here, we illustrate the high-order DM CV-QKD protocol with Prepare-and-Measure (PM) scheme. Alice picks a random bit string, with the defined modulation order $M$ and probability distribution $P_{x_k,p_k}$, the symbol string $A_s$ can be obtained, where $k$ is a natural number with values from 1 to $\sqrt{M}$. $P_{x_k,p_k}$ follows discrete Maxwell-Boltzmann distribution (a special Gaussian distribution) as shown in Eq. (1).

$$P_{x_k,p_k} = \frac{\exp(-\nu(x_k^2 + p_k^2))}{\sum_{x_k,p_k} \exp(-\nu(x_k^2 + p_k^2))}, \quad (1)$$

where $\nu$ is a free parameter and $\nu \geq 0$. Notably, the constant composition distribution matcher (CCDM) algorithm is applied for the symbol generation [20]. Then, by separating the symbol string $A_s$ into $x$ and $p$ quadrature, and utilizing the quadrature amplitude modulation function, the discrete modulation signal can be obtained. Thus, the coherent states $|\alpha_k\rangle = |x_k + ip_k\rangle$ can be prepared from the discrete modulation signal. As an illustration, the constellations of DG 64QAM and 256QAM are shown in Fig. 1, which determines the corresponding distribution of quantum states. In Fig. 1, colors indicate the probabilities corresponding to each coherent state. Alice sends $|\alpha_k\rangle$ through the untrusted fiber channel, and Bob measures them with heterodyne detection. After quantum signal recovery and parameters estimation, according to Ref. [15], the asymptotic secret key rates with a realistic receiver model can be calculated.

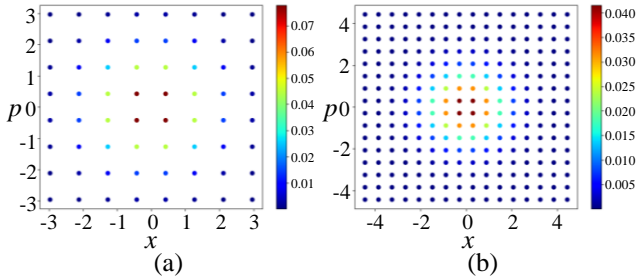

Fig. 1 Constellations of (a) DG 64QAM with $\nu$ = 0.07 and (b) DG 256QAM with $\nu$ = 0.035.

To achieve maximum SKR, before the experiments, simulations have been performed for optimization of key parameters (i.e., free parameter $\nu$ and modulation variance $V_A$), and the results under different distances are shown in Fig. 2. Considering the variation of excess noise under different distances, the excess noise $\xi$ is set to 0.02, 0.05 and 0.08 in the simulations, respectively. As can be seen from Fig. 2, the major trends of DG 64QAM and 256QAM are similar, which show the optimal modulation variance $V_A$ decreases with the secure transmission distance and optimal $\nu$ increases with the secure transmission distance. Furthermore, the influence of excess noise is limited in choosing the optimal value of $\nu$ and $V_A$. In the case of a 25 km transmission link, for example, the optimal value of $\nu$ and $V_A$ are ~0.08 and 4 for DG 64QAM, and ~0.04 and 6 for DG 256QAM, respectively. Therefore, the optimal values of $\nu$ and $V_A$ under different transmission distances can be obtained for experimental demonstrations.

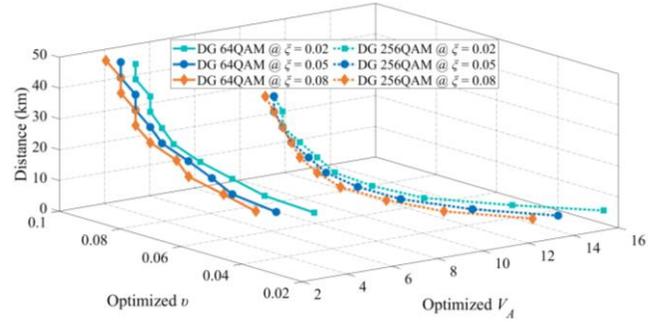

Fig. 2 The simulation results of optimized $\nu$ and $V_A$ for different values of distance and excess noise to achieve maximum SKR.

Figure 3 shows the experimental schematic of the LLO DM CV-QKD system. At Alice's site, a continuous-wave laser (i.e., CW laser 1) with a linewidth of <100 Hz is used as the carrier, and the wavelength is set to 1550.22 nm. Then, the light is split into two branches by a polarization beam splitter (PBS). One branch of the optical carrier is modulated by an In-phase/quadrature modulator (IQ modulator). The $x$ and $p$ quadrature signals with a 750 MHz frequency shift are generated by a two channels arbitrary waveform generator (AWG) that works at 30 GSa/s, and the two electrical signals are amplified by two amplifiers for driving the IQ modulator generating discrete modulation signal. Here, the symbol rate is set to 1 Gbaud, and a digital root-raised cosine (RRC) pulse shape filter with a roof-off factor of 0.3 is performed. Notably, the total block size is set to $2^{22}$. Then, a variable optical attenuator (VOA) is used to adjust the modulation variance $V_A$, and the optical discrete modulation coherent state (DMCS) signal can be achieved. The state of polarization (SOP) of the optical DMCS signal is controlled by a polarization controller (PC) and aligns to the principal axis of the polarization beam combiner (PBC). The reference path mainly consists of an optical delay line (ensure the reference signal has a similar phase characteristic to the optical DMCS signal), VOA (control the optical power of the reference signal), and PC (align the SOP of the reference signal to the other principal axis of PBC). Therefore, the optical DMCS and high-power reference signal are multiplexed by the dimension of polarization and frequency, as shown in the inset of Fig. 3. In the transmission link, SSMF is used. At Bob's site, another independent CW laser with a linewidth of <100 Hz is used as the local oscillator, and the center wavelength is about 1.5 GHz shift from CW laser 1. Then, the optical signal and local oscillator are coherently detected by a polarization diversity receiver. The polarization diversity receiver consists of two PBS, two polarization-maintaining optical couplers (PMOC), and two balanced photo-detectors (BPD). The bandwidth, responsibility, and gain of the BPD are 1.6 GHz, 0.85 A/W, and 1.6×10⁴ V/A, respectively. Finally, the received electrical signals are digitalized by a digital storage oscilloscope (DSO) at 10 Gsa/s, and offline DSP is performed for signal demodulation and parameters estimation.

The main DSP algorithms are: 1) Bandpass filtering and frequency shift. A frequency-domain ideal filter is used for quantum and reference signals, and the bandwidth is 1.3 GHz and 10 MHz, respectively. Meanwhile, a 750 MHz digital frequency shift is performed for the reference signal to remove the frequency difference between the quantum and reference signal. 2) Digital demodulation and carrier recovery. The *x* and *p* quadrature of quantum states are demodulated from the intermediate frequency signal digitally, and carrier frequency shift and phase noise introduced by Alice's and Bob's laser is compensated with the help of the high-power reference signal. 3) Matched filtering and pilot-aided equalization. As the matched filter, an RRC filter with a roll-off factor of 0.3 is used to filter the baseband quantum signal. Then, with the help of the training sequence and least-mean-square (LMS) algorithm, a real-valued finite-impulse response (FIR) filter is implemented. Here, the main functions of the real-valued FIR filter are polarization variation correction, *x* and *p* quadrature imbalance compensation, and residual inter-symbol interference removal. 4) Parameters estimation. Finally, to evaluate the SKR performance, parameters estimation is performed.

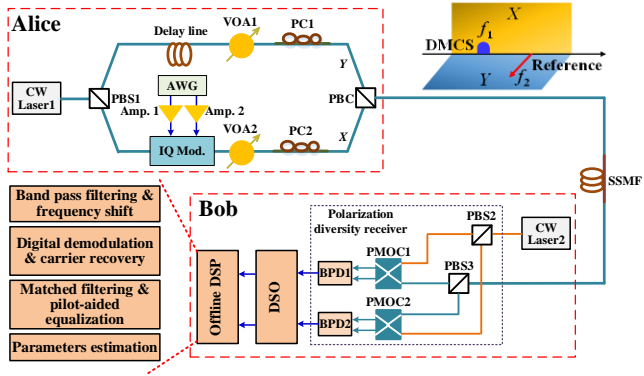

Fig. 3 The experimental schematic of discrete modulated LLO CV-QKD. CW, continuous-wave; PBS, polarization beam splitter; AWG, arbitrary waveform generator; Amp., amplifier; VOA, variable optical attenuator; IQ Mod., In-phase/quadrature modulator; PC, polarization controller; PBC, polarization beam combiner; DMCS, discrete modulated coherent-state; SSMF, standard single-mode fiber; PMOC, polarization-maintaining optical coupler; BPD, balanced photo-detector; DSO, digital storage oscilloscope; DSP, digital signal processing.

**Table 1. Key parameters and measured averaged results for DG 64QAM and 256QAM**

| Modulation format | Distance (km) | $\nu$ | $V_A$ (SNU) | $\xi_A$ (SNU) | SKR (Mbps) |
|---|---|---|---|---|---|
| DG 64QAM | 5 | 0.057 | 7.618 | 0.020 | 288.421 |
| | 10 | 0.064 | 6.513 | 0.023 | 159.395 |
| | 25 | 0.079 | 4.457 | 0.022 | 50.004 |
| | 50 | 0.086 | 3.967 | 0.042 | 7.579 |
| DG 256QAM | 5 | 0.023 | 14.350 | 0.037 | 326.708 |
| | 10 | 0.027 | 12.319 | 0.032 | 179.348 |
| | 25 | 0.039 | 6.332 | 0.029 | 50.623 |
| | 50 | 0.046 | 4.030 | 0.042 | 9.212 |

The key parameters and measured averaged results for DG 64QAM and 256QAM are shown in Table 1. The quantum efficiency of the receiver $\eta = 0.56$, the electronic noise $V_{ele}$ is around 0.15 in the shot noise unit (SNU), reconciliation efficiency $\beta = 0.95$, and the ratio of training sequence (i.e., $P_{TS}$) is set to 20%. In this case, the formula for the asymptotic SKR is given as [15]

$$SKR = R_S(1-P_{TS})(\beta I_{AB} - \chi_{EB}), \qquad (2)$$

where $R_S$ is the Baud rate of the system, $I_{AB}$ is the mutual information between Alice and Bob, and $\chi_{EB}$ is the Holevo information between Eve and Bob. Meanwhile, the influence of $\eta$ and $V_{ele}$ are considered with a realistic model for the receiver [21]. The measured excess noise at Alice's site at transmission distances of 5 km, 10 km, 25 km, and 50 km are shown in Fig. 4, respectively. Here, 20 tests have been made at each transmission distance, and the interval between each test is about 3 minutes. Notably, the excess noise is estimated using the linear channel model with a block size of 1×10⁶. The scattered points in Fig. 4 represent the measured excess noise, and the stars represent the averaged value. A similar characteristic of the excess noise can be seen in Table 1 and Fig. 4 at 5 km, 10 km, and 25 km. However, the fluctuation of excess noise becomes obvious at 50 km. This is mainly because the link becomes more complex and the influence of shot noise fluctuation began to grow. Furthermore, the optimal $V_A$ of DG 256QAM is much larger than DG 64QAM at the short-reach distance, which leads to higher excess noise. However, DG 256QAM is still maintaining a better SKR. It means that the larger QAM has a better excess noise tolerance.

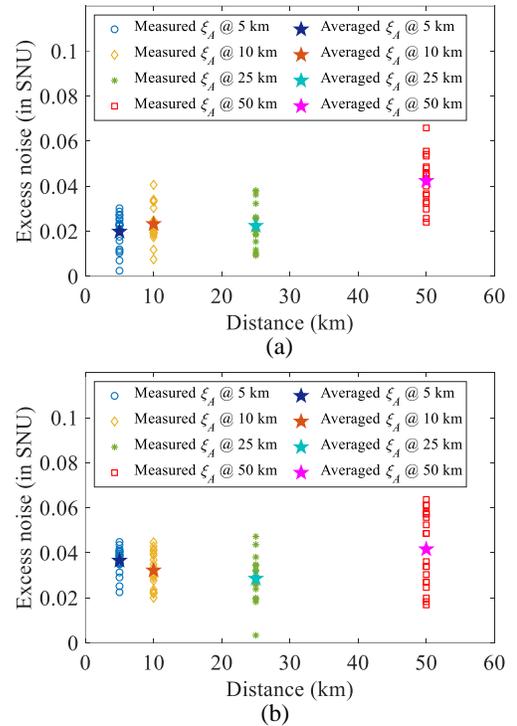

Fig. 4 The excess noise performance under different distances for (a) DG 64QAM, and (b) DG 256QAM.

Figure 5 shows the experimental results of achieved SKR at different distances. In Fig. 5, the curves represent the asymptotic SKR bound by utilizing the parameters in Table 1 (ie. Para. 1/5, 2/6, 3/7, and 4/8 represent the parameters used at the transmission distances of 5 km, 10 km, 25 km and 50km for DG 64QAM/256QAM, respectively), and the circle and diamond dots represent the experimental measured averaged SKR for DG 64QAM and 256QAM, respectively. From Table 1, the asymptotic SKRs are 288.421/326.708 Mbps, 159.395/179.348 Mbps, 50.004/50.623 Mbps and 7.579/9.212 Mbps at the transmission distances of 5 km, 10 km, 25 km and 50km for DG 64QAM/256QAM, respectively. As can be seen from Fig. 5, compared to theoretical calculated SKR curves, a better SKR performance can be achieved by utilizing the optimized parameters at 5 km, 10 km and 25 km. However, the SKR is lower than the theoretical curve (compared to the curve of 25 km) at 50 km, especially in the case of DG 64QAM. The main reason is the excess noise fluctuation becomes worse at a longer transmission distance, which affects the SKR significantly. Furthermore, as a reference, the SKRs of ideal Gaussian modulation are calculated with the same parameters in Table 1. Fig. 5(a) shows that the performance of DG 64QAM is close to the ideal Gaussian modulation, and the performance difference between DG 256QAM and ideal Gaussian modulation is negligible as shown in Fig. 5(b). Meanwhile, some representative experimental works are given in Fig. 5. Compared with the reported CV-QKD experiments, a significant improvement of SKRs can be observed. Consequently, the researched LLO CV-QKD system has a practical potential for high-speed short-reach secure key distribution.

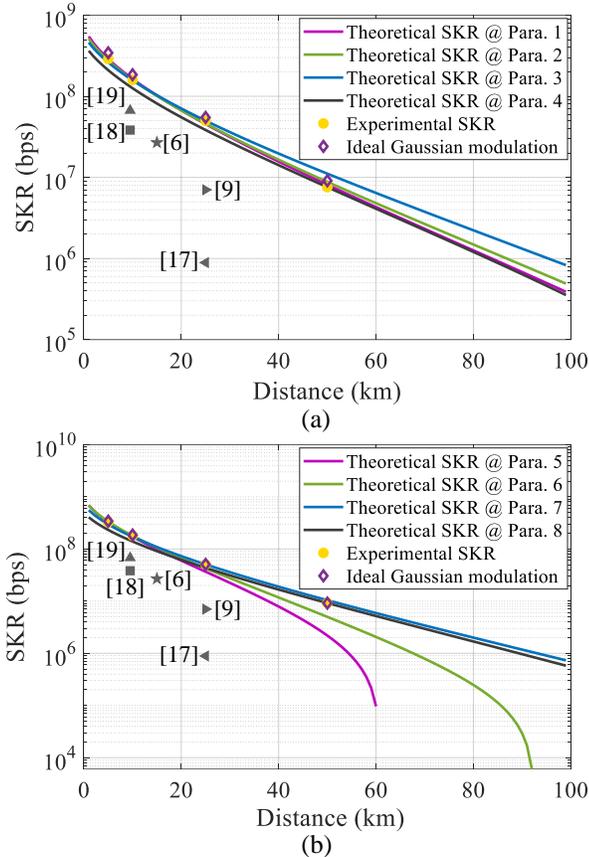

Fig. 5 SKRs vs. secure transmission distances. (a) DG 64QAM, and (b) DG 256QAM. The five grey points represent the experimental SKR value obtained from the corresponding references. Para., parameter. Para. 1/5, 2/6, 3/7, and 4/8 represent the parameters shown in Table 1 at the transmission distances of 5 km, 10 km, 25 km and 50km for DG 64QAM/256QAM, respectively.

Through the optimized key parameters (i.e., $v$ and $V_A$), novel system architecture, and efficient DSP algorithms, here, we experimentally investigated the high order discrete modulated LLO CV-QKD system to achieve a higher SKR and longer transmission distance. The system performances of DG 64QAM and 256QAM have been analyzed. With the optimized parameters, at the transmission distance of 5 km, 10 km, 25 km and 50km, the experimental results show that the asymptotic SKRs are 288.421 Mbps/326.708 Mbps, 159.395 Mbps/179.348 Mbps, 50.004 Mbps/50.623 Mbps and 7.579 Mbps/9.212 Mbps for DG 64QAM/256QAM, respectively. It is notable that, the SKR will be decreased significantly if the finite-size effect is considered. However, especially for tens of kilometers QKD system, the block size of post-processing can be increased and novel post-processing algorithms will be developed to reduce the impact of the finite-size effect. Thus, it can be expected that the demonstrated LLO CV-QKD could be a favorable candidate for short-reach secret communication.

**Funding.** The National Science Foundation of China (Grants No. 62101516, No. 62171418, No. U19A2076, and No. 61901425), the Chengdu Major Science and Technology Innovation Program (Grant No. 2021-YF08-00040-GX), the Technology Innovation and Development Foundation of China Cyber Security (Grant No. JSCX2021JC001), the Sichuan Application and Basic Research Funds (Grant No. 2021YJ0313), the Sichuan Science and Technology Program (Grants No. 2019JDJ0060 and No. 2020YFG0289), National Science Key Lab Fund Project (Grant No. 6142103200105).

**Disclosures.** The authors declare no conflicts of interest.

**Data availability.** Data underlying the results presented in this paper are not publicly available at this time but may be obtained from the authors upon reasonable request.